\documentclass{article}
\usepackage{graphicx}
\usepackage{geometry}
\begin{document}

\title{Critical Threshold For SIRS Model on Small World Networks} 
\author{M. Ali Saif\\
Department of Physics\\
Faculty of Education
University of Amran
Amran,Yemen\\
masali73@gmail.com}

\maketitle
\begin{abstract}
We study the phase transition from the persistence phase to the extinction phase for the SIRS (susceptible/ infected/ refractory/ susceptible) model of diseases spreading on small world network. We show the effects of all the parameters associated with this model on small world network and we create the full phase space. The results we obtained are consistent with those obtained in Ref.\cite{kuperman} in terms of the existence of a phase transition from a fluctuating endemic state to self-sustained oscillations in the size of the infected subpopulation at a finite
value of the disorder of the network. And also our results assert that, that transition specifically occurs where the average clusterization shifts from high to low. The effect of clustering coefficient on SIRS model on the networks can be understood from the results obtained in Ref. \cite{ali}, which indicates the importance of existing the loops in the network, in order to the disease to spread frequently throughout the nodes of network. where, clusters tend to spread infection among close-knit neighborhoods. Hence, when the loops are high inside the network, the reinfection occurs in the network at many places and at different times, which looks like as a kind of randomness in occurring the second period of infection. Whereas when the number of loops are low, reinfection occurs at specific places and times on the network, which looks like as a kind of regularity in occurring the second period of infection.


\end{abstract}
      

\section{Introduction}
In the modeling of many interacting particles on the networks, 
the effect of the networks structure on the properties of dynamical systems
defined on such networks has been attracted a lot of attention recently, and researchers from
fields ranging from neurodynamics and ecology to social sciences have been
extensively working in this area \cite{alb,new,bar,man,lew,watts,kuperman,gade}.
In small world networks \cite{watts}, one starts with a ring of $N$ nodes, in which each node connected to its $k$ nearest neighbors on either side. Then each link from a site to its nearest neighbor is reconnected to another randomly chosen lattice site with probability $p$. This model is proposed to mimic real life situations in which non-local connections exist along with predominantly local.

Kuperman and Abramson \cite{kuperman} studied SIRS model on small world network with the following assumption:
 The susceptible node at time $t$, will be infected at time $t + 1$ with probability proportional
to the fraction of infected nodes in its neighborhood. In other words, if
$\tau_i(t) = 0$, then $\tau_i(t+1) = 1$ with the probability $\lambda_i = k_{inf} /k_i$ where $k_i$ are total number
of neighbors of site $i$, of which $k_{inf}$ are infected. With probability $1 - \lambda_i$,
susceptible node does not change state. The dynamics for the infected node is
deterministic. The infected node slowly become refractory and then eventually
become susceptible again. For the values of $k=3$, $\tau_I=4$ and $\tau_R=9$, they found that, for the more ordered systems, there is a fluctuating endemic state
of low infection. However, at a finite value of the disorder of the network, they get a transition to self-sustained oscillations in the size of the infected subpopulation.
In this work we illustrate the effect of all parameters associated with this system on small world network.

\subsection{Simulation Results}

Here we study the effect of infection time $\tau_I$ on the steady state of the original model of Kuperman and Abramson \cite{kuperman}, as the long rang connection $p$ is changed, for the case when $\tau_R>\tau_I$. We set the values of other parameters as that in original model unless we state different.  

Fig. 1, shows the effect of the increasing on the value of the infection time on the average value of the density of infected nodes $n_{inf}(t)$ of this model on regular a one dimensional lattice. In that figure, we plot the density of infected nodes as function of time at different values of $\tau_I$ and $\tau_R$. For each curve on the figure we set the value of $\Delta \tau=\tau_R-\tau_I$ to be minimum, i. e. $\Delta \tau=1$. As the figure shows, the density of the infected nodes during the first infection period increases as the value of $\tau_I$ increases and that density reaches the maximum value when $\tau_I=10$.

\begin{figure}[htb]
 \includegraphics[width=70mm,height=60mm]{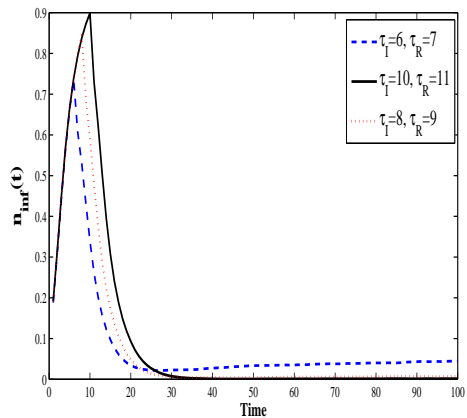}
\caption{(color on line) Density of infectious nodes as function of time for different values of $\tau_I$ and $\tau_R$, as shown in the legend. Other parameters are $N=10^4$, $k=3$, $p=0.0$ and $n_{inf}(0)=0.1$.}
 \end{figure}
 
It is clear from Fig.1 that, when $\tau_I=10$,  and after a short time, all the nodes on the network become sick (where the density of infected nodes initially is $n_{inf}(0)=0.10$, and after 11 time-steps it becomes $n_{inf}(11)=0.90$) during the first infection period, hence system goes to the infection free state. For this case it is evident that, any infectious node on the network, infects all of its neighbors during its first infection period, so according to Ref. \cite{ali} this system reaches extinction state, where all the nodes on the network become susceptible, and also the probability of getting two neighbors which were being infected with time difference $ t >\tau_R$ will be zero (see Fig. 2, when $\tau_I=10$). Therefor, for this case all the nodes on the network pass only through one infection period, and system goes to an absorbing state.  

However, for the case when $\tau_I=6$ and $\tau_I=8$ as Fig. 1 shows, the density of infected nodes approaches the maximum value during the first infection period, while there are a significant numbers of nodes still unaffected. That means on the average, each infected node on the network does not infect all of its neighbors during its infection period. Hence, those uninfected nodes previously, there is a possibility to become lately infected by their second or third etc. infected neighbor. Thus in this case, the probability to get a two neighbors on the network with $ t >\tau_R$ is possible (see Fig. 2, when $\tau_I=8$ and $6$). This behavior prevents the system from falling to an absorbing state from the first infection stage \cite{ali}. 

For the same values of parameters in Fig. 1, we represent in Fig. 2 the density of pairs of neighbors which they have been infected with time difference $t>\tau_R$, as function of time. In calculation  that density, we consider only the nodes in the states $I$ and $R$. It is clear that, the density of pairs of neighbors which they have been infected with time difference $ t>\tau_R$ decreases as the value of $\tau_I$ increases. Figure shows that, when  $\tau_I=10$, the density of pairs of neighbors with $t>\tau_R$ goes to zero. However, when $\tau_I<10$ there are significant numbers of pairs of neighbors with $t>\tau_R$.

\begin{figure}[htb]
\includegraphics[width=70mm,height=60mm]{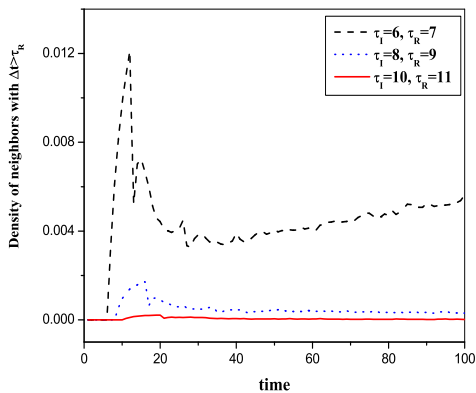}
\caption{ (color on line)Density of pairs of nodes have $t>\tau_R$ as function of time for different values of $\tau_I$ and $\tau_R$, as shown in the legend. Other parameters are $N=10^4$, $k=3$, $p=0.0$ and $n_{inf}(0)=0.1$.}
 \end{figure}

For completeness, we examine the model when the value of $\Delta\tau$ increases at various values of $\tau_I$ and $\tau_R$. We find that, when $\tau_I=9$, system goes to the extinction state when $\tau_R=12$. This corresponds to $\Delta \tau=3$. In general we find that, for any values of $\tau_I$ and $\tau_R$ which satisfy the condition $\tau_I<\tau_R$, the system evolves to an extinction when $\tau_I+\tau_R=21$.

 Situation becomes more complicated on the small world network where, the nodes have different numbers of nearest neighbors $k_i$. There are nodes become heavily connected, such nodes will need less time on the average until they become infected. However, there are some other nodes become less connected which means on the average they will need longer time until they become infected. We have performed extensive numerical simulations at different values of $p$ ranging from $\left[0.01-1.0\right]$. Interestingly we find that, for any value of $p$, the system reaches an extinction state when $\tau_I+\tau_R \approx 21$, in behavior similar to what happens on the regular lattice. This result is expected where, small world network of Watts and Strogatz which we use in our network has on average a fixed connectivity $\left\langle k\right\rangle=2k$ \cite{watts} for any values of the disorder parameter $p$. 

Finally, we study the effects of the parameters $\tau_I$, $\tau_R$ and $p$ on the steady state of this model. Fig. 3 shows, the density of infected nodes as function of time at different values of $\tau_I$ and $\tau_R$. Fig. 3a, shows three time series of $n_{inf}(t)$ when the value of the disorder parameter $p$ is $p=0.2$. In these curves, we fix the value of $\tau_I=6$, and $\tau_R$ takes the values $\tau_R=9$ (bottom), $11$ (middle), and $12$ (top). It is evident that, as the value of $\tau_R$ increases, the system crosses from the fluctuating endemic state (when $\tau_I=9$) to an oscillatory state (when $\tau_I=11$). Even if the amplitude of oscillation is slightly small, but it is almost periodic with a very well defined period. Fig. 3b shows two time series of $n_{inf}(t)$ when the value of the disorder parameter $p$ is $p=0.8$. In the two curves, we fix the value of $\tau_I=6$, and $\tau_R$ takes the values $\tau_R=7$ (top), and $10$ (bottom). It is clear in this case, the large amplitude self-sustained oscillation is developed.

 \begin{figure}[htb]
 \includegraphics[width=70mm,height=60mm]{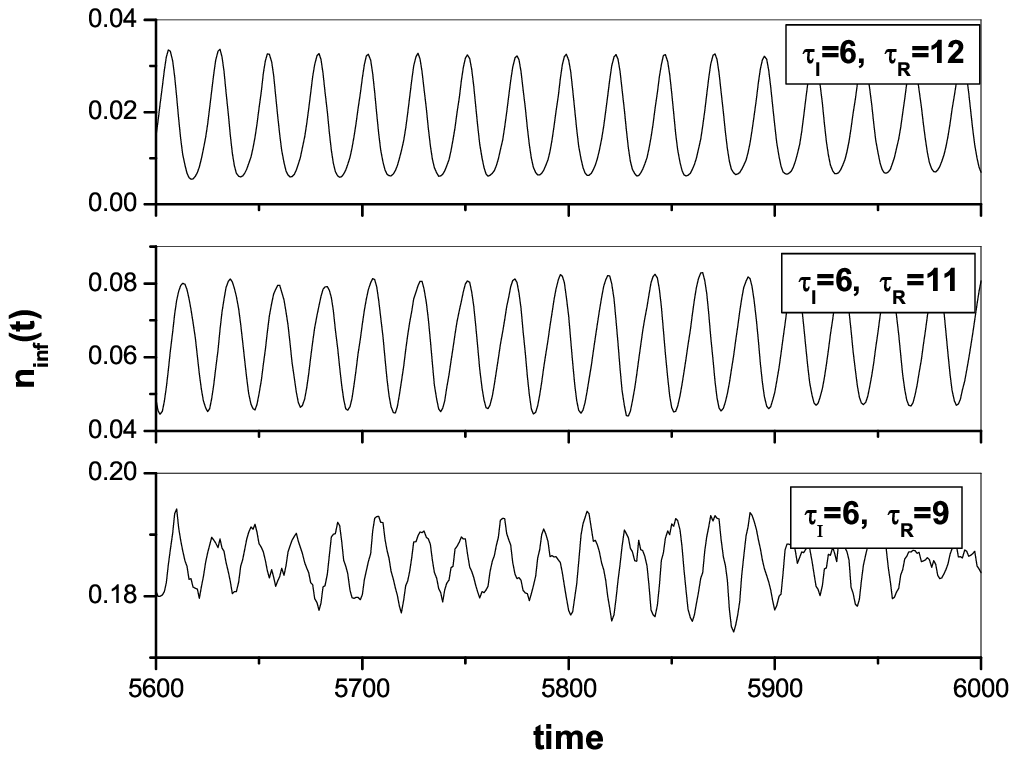}
 \includegraphics[width=70mm,height=60mm]{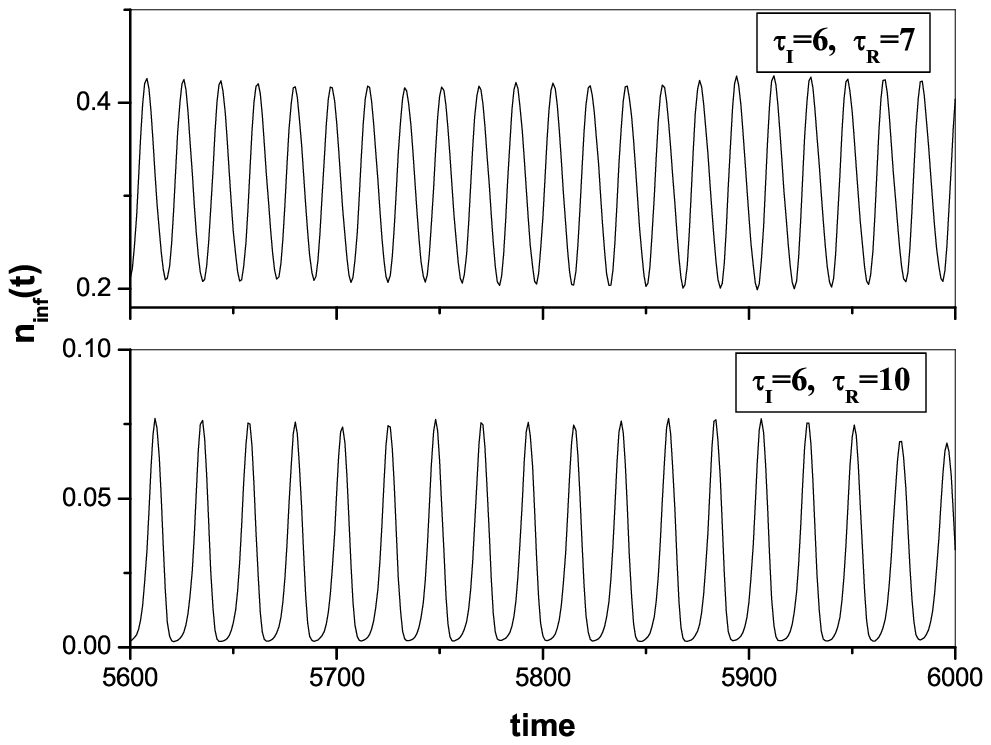}
 \caption{Fraction of infected nodes as a function of time for the last of $6000$ time steps. (a) Three time series are shown corresponding to different values of $\tau_I$ and $\tau_R$, for $p=0.2$. (b) Two time series are shown corresponding to different values of $\tau_I$ and $\tau_R$, for $p=0.8$. All the curves have $N=10^4$, $k=3$ and $n_{inf}(0)=0.1$. Each curve averaged over $20$ configurations.}
 \end{figure}

In Fig. 4, we create the phase space of the SIRS model at several values of $p$. For each value of $p$, we study the system at various values of $\tau_I$ and $\tau_R$. We find that, when value of the disorder parameter $p$ is bigger than $0.14$, we can distinguish between three phases: a susceptible-absorbing phase, a self-sustained oscillation phase and a fluctuating endemic phase. Whereas, when $p< 0.14$ we observe only two phases, a susceptible-absorbing phase and a fluctuating endemic phase. 

In Fig. 4, for the case when $p=0.1$ the regions II+III+IV (the regions under the black solid line) are corresponding to the fluctuating endemic phase, whereas the region I (the region upper the black solid line) is corresponding to an absorbing phase. The black solid line is the critical line that separates the absorbing phase from the coexiectence stable phase. However when $p=0.2$, the model shows the three phases, the susceptible-absorbing phase is the region I (the region upper the black solid line), self-sustained oscillation phase is corresponding to the region II (the region enclosed by the solid line and the dotted curve), and the regions III+IV are corresponding to the fluctuating endemic phase (the remaining region under the solid line and the dotted curve). The dotted curve is the critical curve that separates the oscillation phase from the fluctuating endemic phase. When, $p=0.8$ 
the oscillation phase is corresponding to the regions II and III (the region enclosed by the solid line and the dashed curve). It is clear that, the region corresponding to the oscillation phase shrinks as the value of $p$ decreases and becomes wider as the value of $p$ increases. Here, we can infer that, the critical value of $p$, which separates the oscillation phase from the fluctuating endemic phase, should be in between $0.1<p_c<0.2$. For best estimate, the critical point is approaching the value $p_c=0.14\pm0.02$, when $\tau_I=7$ and $\tau_R=13$. 

\begin{figure}[htb]
 \includegraphics[width=70mm,height=60mm]{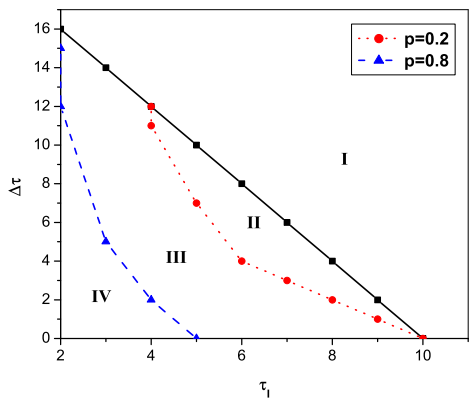}
\caption{(color on line)Phase diagram in the $(\tau_I,\Delta \tau)$ plane of our model for values of $p=0.1,0.2,$ and $0.8$. When $N=10^4$, $k=3$ and $n_{inf}(0)=0.1$.
In the case of $p=0.1$, there are only two phase: the region I is corresponding to susceptible-absorbing phase and regions II+III+IV are corresponding to active phase with nonzero infective densities. The critical line between these two phases is the black solid line. For $p=0.2$, there are three phases: the region I is the absorbing phase, the region II is the self-sustained oscillation phase and the regions III+IV are the active phase. The dotted curve is the critical curve separates the oscillation phase from the active phase. For $p=0.8$, there are also three phases: the region I is the absorbing phase, the regions II+III are the self-sustained oscillation phase and IV is the active phase. For $p=0.8$, the dashed curve is the critical curve separates the oscillation phase from the active phase.}
 \end{figure}

The value of $p_c$ we find here approximately is the value of $p$ where the average clusteriztion shifts from high to low as mentioned in Ref. \cite{kuperman}. We support that conclusion with the following argument. It had been proved in Ref. \cite {ali} that, the clustering coefficient will play an important role in the SIRS model, where existence the loops on the network is necessary in order to the disease to spread frequently throughout the nodes of
the networks. Whereas, clusters tend to spread infection among close-knit neighborhoods \cite{lew}. We speculate that, whenever the value of clustering coefficient is high the next period of infection will happen at many places on the network and at any time, which will look like as a kind of randomness (in space and time) in the next generation of infection. However, when the clustering coefficient becomes lower, which means the number of triangular loops on the network also will become lower, the reinfection will be localized where those loops exist, consequently the next period of infection (on the average) will happen at specific place and time on the network. This behavior becomes more apparent as the value of clustering coefficient becomes smaller at higher values of $p$, where the periodicity of oscillation becomes smoother.  

Here, we point out to that, Phase transition at specified randomness values of small world network has been observed also in many systems such as, a propagation of a rumor on small world networks \cite{zan}, and in a system of coupled oscillatory elements, the introduction of shortcuts enhances the network synchronizability \cite{bar1}, also in a self-sustained activity of excitable neurons, the introduction of shortcuts changes the probability of failure from 0 to 1 over a narrow range in $p$ \cite{rox}. In Ising model, the addition of shortcuts induces a finite-temperature phase transition even in the one-dimensional Ising model \cite{pek}, and the introduction of unidirectional shortcuts can change the second-order phase transition in the two dimensional Ising model into a first-order one \cite{san}

 \section{Conclusion} 
We have studied the spreading of infectious diseases for the SIRS model on small world network. We examine the effects of all parameters related to this model on its steady state. we find that, when the disorder parameter is $p>0.14$, we can distinguish between three phases: a susceptible-absorbing phase, a self-sustained oscillation phase and a fluctuating endemic phase. However when $p<0.14$ we find only two phases:  a susceptible-absorbing phase and a fluctuating endemic phase. For best estimate, $p=0.14\pm0.2$ is the critical value which separates the oscillation phase from the fluctuating endemic phase for this model on small world network.

 \end{document}